\documentclass[10pt,runningheads]{llncs}

\usepackage{float}
\usepackage[
    letterpaper,
    margin=1in
]{geometry}

\usepackage{booktabs}
\usepackage[T1]{fontenc}
\usepackage{newtxtext}
\usepackage[varvw]{newtxmath}
\usepackage{graphicx}
\usepackage[T1]{fontenc}
\makeatletter
\AtBeginDocument{%
    \let\small\normalsize

}
\makeatother
\usepackage{newtxtext}       % 
\usepackage[varvw]{newtxmath}       % selects Times Roman as basic font

\usepackage{graphicx}
\begin{document}
%
% \title{EvoSurv: Evolution-Aware Multimodal Survival Modeling for Assessing the Prognostic Value of Tumor Clonality}
% \title{EvoSurv: Evolution-Aware Multimodal Survival Modeling Reveals the Prognostic Value of Tumor Evolution}
\title{EvoSurv: Evolution-Aware Multimodal Survival Modeling Reveals Complementary Prognostic Signals}

\author{
Wei Wang\inst{1} \and
Po-Yu Liang\inst{2} \and
Jun Bai\inst{1}\thanks{Corresponding author.}
}

\authorrunning{W. Wang et al.}

\institute{
Department of Computer Science,
University of Cincinnati,\\
Cincinnati, OH 45221, USA\\
\email{wang3wa@mail.uc.edu, baiju@ucmail.uc.edu}
\and
Department of Microbiology and Plant Pathology,\\
University of California, Riverside,\\
Riverside, CA 92521, USA\\
\email{poyul@ucr.edu}
}

% \author{Wei Wang\inst{1}\orcidID{0000-1111-2222-3333} \and
% Po-Yu Liang\inst{2,3}\orcidID{1111-2222-3333-4444} \and
% Jun Bai\inst{3}\orcidID{2222--3333-4444-5555}}
% %
% \authorrunning{W. Wang et al.}
% Springer Heidelberg, Europaplatz 3, 69115 Heidelberg, Germany
% ABC Institute, Ruprecht-Karls-University Heidelberg, Heidelberg, Germany\\
\maketitle              % typeset the header of the contribution
\begin{abstract}

Accurate cancer survival prediction remains challenging due to tumor heterogeneity. Although multimodal survival models integrate clinical and molecular data, they often overlook evolutionary information encoded in genomic alterations. To address this gap, we developed eTime, a 38-dimensional evolutionary representation comprising eight feature groups that characterize clonal and subclonal mutation burden, mutational spectra, functional consequences, copy-number alterations, and anoikis-related genomic changes. We further developed EvoSurv, a multimodal survival framework integrating eTime with clinical variables and RNA expression.
Across six TCGA cancer cohorts comprising 2,782 patients, EvoSurv achieved the highest mean concordance index (C-index) in all six cancers, averaging 0.663 and outperforming three state-of-the-art survival prediction methods. 
More importantly, we assessed whether evolutionary features provide additional prognostic information beyond clinical and transcriptomic predictors. eTime remained significantly associated with survival in head and neck squamous cell carcinoma (HNSC) and liver hepatocellular carcinoma (LIHC) after adjustment for clinical variables and RNA-derived risk scores. Feature-group analyses further revealed cancer-specific prognostic patterns. In HNSC, T>G substitutions were associated with increased mortality risk after clinical and molecular adjustment, including in HPV-negative patients, with supporting evidence from an independent CPTAC-HNSCC cohort.
Overall, these findings demonstrate that eTime captures survival-associated evolutionary information and highlight its potential to provide prognostic value beyond clinical and transcriptomic predictors. 
% The source code is available at \url{https://github.com/LabJunBMI/EvoSurv.git}.


\end{abstract}

\newpage
\section{Introduction}

Cancer development and progression are driven by continuous somatic evolution. Through mutation accumulation and clonal selection, tumors develop genetically distinct clonal and subclonal populations, giving rise to extensive intratumor heterogeneity~\cite{black2021genetic,Dentro2021}. These genomic alterations reflect tumor evolutionary history and may contribute to differences in disease progression and patient outcomes. Patients with similar clinical characteristics can exhibit different survival outcomes~\cite{van2002gene}, highlighting the potential relevance of tumor molecular heterogeneity to prognosis. Understanding the prognostic significance of tumor evolutionary features may therefore provide further insight into the molecular basis of survival heterogeneity.

However, whole-exome sequencing typically captures the genomic state of a tumor at a single time point, limiting direct observation of its evolutionary history. Nevertheless, mutation-level cancer cell fractions  provide insights into clonal evolution~\cite{black2021genetic,Dentro2021}. Clonal mutations are shared across the sampled cancer cell population and are generally associated with ancestral lineages, whereas subclonal mutations occur in only a subset of cancer cells, reflecting subsequent clonal diversification. This distinction enables the characterization of mutation features across different clonal states from a single sequencing sample. Large-scale cancer genomic studies have further characterized multiple dimensions of tumor evolution, including clonal and subclonal alterations, mutational spectra, and copy-number changes~\cite{Dentro2021,alexandrov2020repertoire}. Studies of specific genomic alterations and evolutionary patterns have also identified associations with patient survival~\cite{bielski2018genome,Fontana2023}.   %However, systematically representing these evolutionary characteristics and assessing their prognostic contributions across cancer types remain challenging. In particular, whether evolutionary features provide complementary prognostic information beyond clinical and transcriptomic predictors remains unclear.

In recent years, advances in deep learning have expanded cancer survival prediction to high-dimensional molecular data and multimodal analysis. DeepSurv~\cite{katzman2018deepsurv} and Cox-nnet~\cite{ching2018cox} employ neural networks to model survival risk, while more recent approaches, including AGGSurv~\cite{AGGSurv}, FGCNSurv~\cite{FGCNSurv}, and CoFormerSurv~\cite{CoFormerSurv}, incorporate molecular information through representation learning and multimodal integration. Despite these advances, clonal and subclonal evolutionary patterns remain underrepresented in existing survival prediction frameworks. Incorporating biologically informed evolutionary features may therefore provide additional insight into the relationship between genomic heterogeneity and patient survival.

To systematically investigate the prognostic value of tumor evolutionary features, we developed eTime, an evolution-aware feature representation, and EvoSurv, a multimodal survival analysis framework. Our main contributions are summarized as follows:

\begin{enumerate}
    \item \textbf{Evolution-aware feature representation.} We developed eTime, a 38-dimensional representation that summarizes clonal and subclonal mutation patterns, predicted functional effects, copy-number alterations, and genomic changes in anoikis-related genes.
    \item \textbf{Multimodal survival modeling.} We developed EvoSurv to integrate evolutionary, transcriptomic, and clinical information for survival prediction. EvoSurv achieved the highest mean concordance index (C-index) among the four evaluated methods in each of the six The Cancer Genome Atlas (TCGA) cohorts.
    \item \textbf{Prognostic assessment and cancer-specific findings.} We compared an eTime-only model with a shuffled-feature baseline and tested whether eTime risk scores remained associated with survival after adjustment for clinical and RNA-derived risk scores. The eTime risk score remained associated with survival after joint adjustment in head and neck squamous cell carcinoma (HNSC) and liver hepatocellular carcinoma (LIHC), while feature-level analyses revealed cancer-specific prognostic patterns. We further examined cancer-specific feature associations and evaluated the HNSC T>G signal in an independent Clinical Proteomic Tumor Analysis Consortium (CPTAC)-HNSCC cohort. 
\end{enumerate}

\vspace{-2em}
\section{ Methods and Materials }

\begin{figure}[h]
    \centering
    \includegraphics[width=\linewidth]{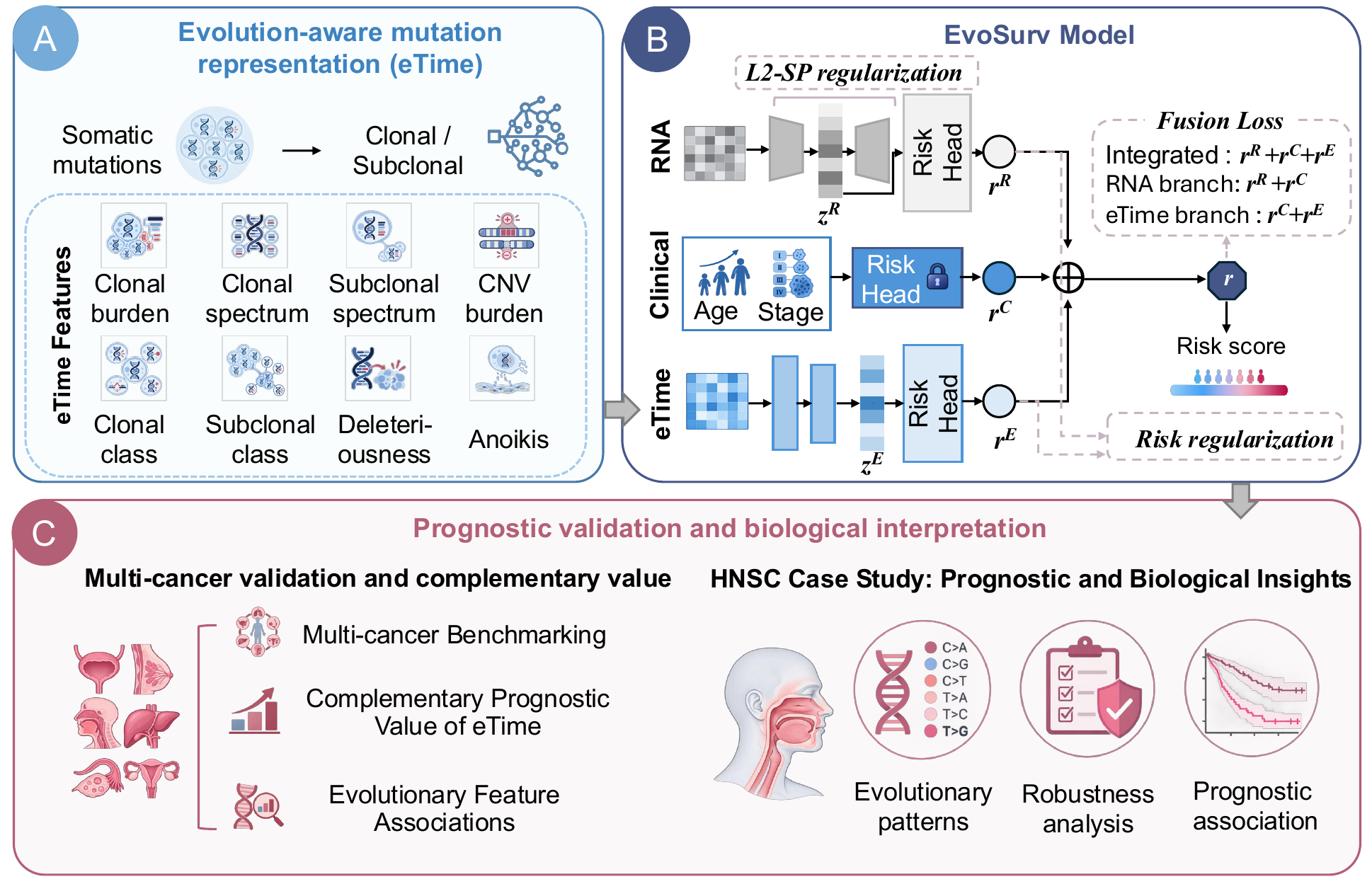}
    \caption{\textbf{Overview of the EvoSurv framework.}
(A) Eight evolutionary feature groups derived from somatic mutations and copy-number alterations.
(B) Additive fusion of RNA, eTime, and fixed clinical risk with joint survival supervision.
(C) Multi-cancer prognostic evaluation and HNSC case study of mutational patterns and robustness.
Some of the icons were generated using generative AI and are for illustrative purposes only.
}
    \label{fig:framework}
    \vspace{-1.5em}
\end{figure}

The EvoSurv framework consists of three components: evolution-aware mutation representation, multimodal survival modeling, and prognostic evaluation (Fig.~\ref{fig:framework}). We describe the construction of eTime and the EvoSurv model, followed by the datasets, experimental setup, and statistical analyses used to evaluate evolutionary features across cancer cohorts.

\vspace{-1em}
\subsection{Evolution-Aware Mutation Representation}
\label{etime}

To characterize tumor evolutionary heterogeneity, we developed eTime, a 38-dimensional representation comprising eight feature groups spanning clonal composition, mutational patterns, functional consequences, and genomic alterations. By distinguishing clonal and subclonal mutations based on their cancer cell fractions (CCFs), eTime captures genomic characteristics associated with different clonal states (Fig.~\ref{fig:framework}A).

Somatic mutation data, including read counts, functional annotations, and SIFT scores, were obtained from  TCGA~\cite{TCGA} through the Genomic Data Commons~\cite{heath2021nci}. Tumor purity and copy-number estimates were obtained from ABSOLUTE~\cite{Absolute}. Mutation read counts, tumor purity, and copy-number information were used to infer CCFs and clonal clusters with PyClone~\cite{PyClone}. For tumor $i$, let $\mathcal{M}_i$ denote its nonsilent somatic single-nucleotide variants (SNVs) and $\phi_m$ the inferred CCF of mutation $m$. Following previous studies~\cite{CCF}, mutations were partitioned into clonal and subclonal sets:
\begin{equation}
\mathcal{M}_i^{\mathrm{c}}=\{m\in\mathcal{M}_i:\phi_m\geq0.9\},\qquad
\mathcal{M}_i^{\mathrm{s}}=\mathcal{M}_i\setminus\mathcal{M}_i^{\mathrm{c}}.
\label{eq:partition}
\end{equation}

% \noindent\textbf{Clonal composition and mutational spectra (17 features).}

% \paragraph{\textbf{Clonal composition and mutational spectra (17 features).}}
\subsubsection{Clonal composition and mutational spectra (17 features)}
\textit{(1) Clonal burden.} %Five features quantified mutation burden and intratumor heterogeneity: 
% We used five features to quantity mutation burden and intratumor heterogenity: the log-transformed clonal,$\log(1+|\mathcal{M}_i^{\mathrm{c}}|)$; the subclonal mutation counts,$\log(1+|\mathcal{M}_i^{\mathrm{s}}|)$; the subclonal fraction $|\mathcal{M}_i^{\mathrm{s}}|/|\mathcal{M}_i|$, the number of subclonal PyClone clusters $K_i^{\mathrm{s}}$, and Shannon diversity $H_i=-\sum_{q=1}^{K_i}\pi_{iq}\log_2\pi_{iq}$, where $K_i$ is the total cluster count and $\pi_{iq}$ is the fraction of mutations in cluster $q$. 
We used five features to quantify mutation burden and intratumor heterogeneity: log-transformed clonal and subclonal mutation counts, $\log(1+|\mathcal{M}_i^{\mathrm{c}}|)$ and $\log(1+|\mathcal{M}_i^{\mathrm{s}}|)$, the subclonal fraction $|\mathcal{M}_i^{\mathrm{s}}|/|\mathcal{M}_i|$, the number of subclonal PyClone clusters $K_i^{\mathrm{s}}$, and Shannon diversity $H_i=-\sum_{q=1}^{K_i}\pi_{iq}\log_2\pi_{iq}$, where $K_i$ is the total cluster count and $\pi_{iq}$ is the fraction of mutations in cluster $q$.
\textit{(2) Clonal and subclonal mutational spectra.} We classified each SNV as clonal or subclonal and assigned it to one of six pyrimidine-referenced substitution classes~\cite{6mutation}, with reverse-complement substitutions (e.g., A>C and T>G) counted as the same class: C>A, C>G, C>T, T>A, T>C, and T>G. For each clonal state $k\in\{\mathrm{c},\mathrm{s}\}$ and substitution class $j$, we calculated $p_{ij}^{k}=n_{ij}^{k}/|\mathcal{M}_i^{k}|$, where $n_{ij}^{k}$ denotes the number of mutations of class $j$ in state $k$ for tumor $i$.
Fractions were set to zero when their denominators were zero.
Each clonal state contributed six features, yielding 12 features capturing substitution composition independent of mutation burden.

% \paragraph{\textbf{Functional consequences and deleteriousness (12 features).}}
\subsubsection{Functional consequences and deleteriousness (12 features)}
\textit{(1) Clonal and subclonal functional classes.} We derived eight features from mutation consequence annotations, with four categories per clonal state: missense, nonsense, splice-related (splice site or splice region), and other nonsilent variants. As with mutational spectra, each feature was calculated as the fraction of mutations belonging to its category within the corresponding clonal state. 
\textit{(2) Deleteriousness.} We derived four features from SIFT scores~\cite{sift}, calculated separately for clonal and subclonal mutations. SIFT scores range from 0 to 1, with lower scores indicating greater predicted functional impact. For each state, we calculated the proportion of deleterious SNVs (SIFT score $\leq0.05$) among all SNVs and the mean SIFT score across mutations with available scores. If no valid SIFT scores were available, we set the mean to 1.

% \vspace{-0.4em}
% \paragraph{\textbf{Copy-number and anoikis-related alterations (9 features).}}
\subsubsection{Copy-number and anoikis-related alterations (9 features)}
\textit{(1) Copy-number variation (CNV) burden.} We derived four features from ABSOLUTE-based gene-level integer copy numbers.  %to quantify genomic copy-number alterations. 
Relative to the diploid state ($\mathrm{CN}=2$), we calculated the proportions of genes with copy-number changes ($\mathrm{CN}\neq2$), amplification ($\mathrm{CN}\geq3$), and deletion ($\mathrm{CN}\leq1$), as well as the mean absolute copy-number deviation, $D_i^{\mathrm{CN}}=N_i^{-1}\sum_{g=1}^{N_i}|\mathrm{CN}_{ig}-2|$, where $N_i$ is the number of genes with valid copy-number estimates for patient $i$. \textit{(2) Anoikis-related alterations.} We constructed five features to characterize genomic alterations associated with anoikis, a form of apoptosis induced by detachment from the extracellular matrix and linked to metastatic progression~\cite{Anoikis}.
We obtained 1,169 anoikis-related genes from GeneCards~\cite{GeneCards} and retained those represented in each cancer cohort’s RNA-seq expression matrix. The features included the proportion of nonsilent SNVs affecting anoikis-related genes among all nonsilent SNVs, the subclonal fraction among anoikis-related SNVs, a binary indicator of at least one anoikis-related SNV, and the proportions of anoikis-related genes with amplification ($\mathrm{CN}\geq3$) or deletion ($\mathrm{CN}\leq1$).

\subsection{EvoSurv Model}
\label{sec:model}

% EvoSurv integrates tumor evolutionary features with transcriptomic and clinical information to assess their contributions to survival prediction (Fig.~\ref{fig:framework}B). The framework consists of modality-specific molecular risk predictors, additive risk fusion with a fixed clinical offset, and joint survival supervision. By optimizing molecular risk scores under the same clinical baseline, EvoSurv enables the evaluation of evolutionary information alongside established clinical and transcriptomic predictors.

EvoSurv integrates tumor evolutionary features with transcriptomic and clinical information to assess their contributions to survival prediction (Fig.~\ref{fig:framework}B). The framework consists of modality-specific risk predictors, additive risk fusion with a fixed clinical offset, and joint survival supervision. By optimizing molecular risk scores under the clinical baseline, EvoSurv enables the evaluation of evolutionary information alongside clinical and transcriptomic predictors.

\vspace{-1em}
\subsubsection{Molecular Risk Prediction}
For patient $i$, let $\mathbf{x}_i^R\in\mathbb{R}^{2000}$ denote the selected RNA expression features and $\mathbf{e}_i\in\mathbb{R}^{38}$ denote the eTime representation described in Section~\ref{etime}. We used modality-specific transformations to obtain latent representations:
\[
\mathbf{z}_i^R = f_R^{enc}(\mathbf{x}_i^R) \in \mathbb{R}^{64},
\qquad
\mathbf{z}_i^E = f_E(\mathbf{e}_i) \in \mathbb{R}^{16}.
\]%The two modalities were processed by separate encoders to accommodate their different feature distributions.

The RNA autoencoder was pretrained using a reconstruction objective and had a symmetric architecture, its encoder \(f_R^{\mathrm{enc}}\) comprised fully connected layers with dimensions 2000--512--256--64. The eTime transformation \(f_E\) was a two-layer multilayer perceptron. Both transformations used layer normalization, GELU activation, and dropout (\(p=0.3\)). Modality-specific linear heads produced prognostic risk scores:

%For RNA expression, an autoencoder (AE) was pretrained using a reconstruction objective. Its encoder $f_R$ comprised three fully connected layers with dimensions 2000--512--256--64, producing a 64-dimensional representation $\mathbf{z}_i^R=f_R(\mathbf{x}_i^R)\in\mathbb{R}^{64}$. The pretrained encoder was initialized with the learned AE weights and subsequently fine-tuned during survival training.

%For eTime, a two-layer multilayer perceptron $f_E$ mapped the 38-dimensional input into a 16-dimensional latent representation, $\mathbf{z}_i^E=f_E(\mathbf{e}_i)\in\mathbb{R}^{16}$. The encoder incorporated layer normalization, GELU activation, and dropout with a rate of 0.3.

%Two modality-specific linear prediction heads transformed the learned representations into scalar prognostic risk scores:

\begin{equation}
r_i^R=\mathbf{w}_R^\top\mathbf{z}_i^R,\qquad
r_i^E=\mathbf{w}_E^\top\mathbf{z}_i^E,
\label{eq:modality_risk}
\end{equation}
where $\mathbf{w}_R\in\mathbb{R}^{64}$ and $\mathbf{w}_E\in\mathbb{R}^{16}$ are trainable risk-head coefficients.
% These scores are the transcriptomic and evolutionary components of the predicted log-relative hazard, respectively.
% These scores represent the transcriptomic and evolutionary components of the predicted logarithmic relative hazard, respectively. %The corresponding scores $r_i^R$ and $r_i^E$ represent the transcriptomic and evolutionary components of the predicted log-relative hazard, respectively.

\vspace{-1em}
\subsubsection{Multimodal Risk Fusion}

We combined the clinical, RNA, and eTime risk scores using additive risk-level fusion. Clinical risk was estimated with a Cox proportional hazards model fitted to age and tumor stage, with the clinical log-relative hazard defined as $r_i^C = \boldsymbol{\beta}_C^\top \mathbf{x}_i^C$, where $\mathbf{x}_i^C$ is the clinical covariate vector for patient $i$ and $\boldsymbol{\beta}_C$ is the corresponding coefficient vector. The clinical model was fitted on the training data, and its coefficients were held fixed during molecular model optimization. The final risk score was
\begin{equation}
r_i = r_i^C + r_i^R + r_i^E,
\end{equation}
where $r_i^R$ and $r_i^E$ denote the RNA and eTime risk scores, respectively. Thus, the molecular scores contribute risk conditional on the clinical component, while their individual contributions remain interpretable.

% We adopted additive risk-level fusion to combine molecular information % without requiring alignment 
% between the RNA and eTime. %latent spaces. 
% % The molecular risk score was defined as $r_i^M=r_i^R+r_i^E$. 
% Clinical risk was estimated using a Cox proportional hazards model fitted on age and tumor stage. %Let $\mathbf{x}_i^C$ denote the clinical covariate vector and $\boldsymbol{\beta}_C$ its estimated Cox regression coefficients. 
% Let \(\mathbf{x}_i^C\) denote the clinical covariate vector for patient \(i\), and let \(\boldsymbol{\beta}_C\) denote the corresponding Cox regression coefficient vector. The clinical log-relative hazard was defined as $r_i^C=\boldsymbol{\beta}_C^\top\mathbf{x}_i^C$. The clinical model was fitted exclusively on the training data, and its coefficients remained fixed during subsequent molecular optimization. The final prognostic risk score was computed as
% \begin{equation}
% r_i=r_i^C+r_i^R+r_i^E.
% \label{eq:final_risk}
% \end{equation}
% % \begin{equation}
% % r_i=r_i^C+r_i^M=r_i^C+r_i^R+r_i^E.
% % \label{eq:final_risk}
% % \end{equation}
% In this formulation, the fixed clinical risk acts as an offset, allowing the RNA and eTime predictors to estimate additional risk contributions conditional on clinical covariates. The additive structure also preserves separate molecular risk components, enabling their contributions to be examined individually.

\vspace{-1em}
\subsubsection{Survival Learning Objective}
The molecular predictors were jointly trained using the Cox proportional hazards objective. For $N$ patients, let $t_i$ denote the observed follow-up time and $\delta_i\in\{0,1\}$ indicate whether death was observed. Given the vector of predicted log-relative hazards $\mathbf{r}=(r_1,\ldots,r_N)$, the negative Cox partial log-likelihood was defined as
\begin{equation}
\mathcal{L}_{\mathrm{Cox}}(\mathbf{r})
=-\sum_{i:\delta_i=1}
\left[
r_i-\log\sum_{j:t_j\geq t_i}\exp(r_j)
\right].
\label{eq:cox}
\end{equation}
To supervise the integrated prediction and both molecular branches, we used a joint hard supervision objective:
\begin{align}
\mathcal{L}_{\mathrm{fusion}}
={}&\mathcal{L}_{\mathrm{Cox}}(\mathbf{r}^C+\mathbf{r}^R+\mathbf{r}^E)\nonumber
+\mathcal{L}_{\mathrm{Cox}}(\mathbf{r}^C+\mathbf{r}^R)
+\mathcal{L}_{\mathrm{Cox}}(\mathbf{r}^C+\mathbf{r}^E),
\label{eq:fusion_loss}
\end{align}
where $\mathbf{r}^C$, $\mathbf{r}^R$, and $\mathbf{r}^E$ denote the clinical, transcriptomic, and evolutionary risk vectors across patients. The first term optimizes the combined prediction, whereas the two auxiliary terms encourage each molecular branch to retain prognostic predictive capacity conditional on the clinical offset. All three loss terms contribute to updating the molecular encoders and risk heads, while the clinical Cox coefficients remain fixed. To stabilize molecular risk estimation and preserve pretrained RNA representations, we incorporated risk regularization and L2-SP regularization into the overall training objective:
\begin{equation}
\mathcal{L}_{\mathrm{total}}
=\mathcal{L}_{\mathrm{fusion}}
+\lambda_r\mathcal{L}_{\mathrm{risk}}
+\lambda_{\mathrm{SP}}
\|\theta_R-\theta_R^{(0)}\|_2^2,
\label{eq:total_loss}
\end{equation}
where $\mathcal{L}_{\mathrm{risk}}=\frac{1}{2N}\sum_{i=1}^{N}[(r_i^R)^2+(r_i^E)^2]$ penalizes large molecular risk scores, while L2-SP regularization constrains the fine-tuned RNA encoder parameters $\theta_R$ toward their pretrained values $\theta_R^{(0)}$. We set $\lambda_r=10^{-3}$ and $\lambda_{\mathrm{SP}}=10^{-2}$.

% To stabilize molecular risk estimation and preserve pretrained transcriptomic representations during fine-tuning, we introduced two regularization terms. The overall objective was
% \begin{equation}
% \mathcal{L}_{\mathrm{total}}
% =\mathcal{L}_{\mathrm{fusion}}
% +\lambda_r\mathcal{L}_{\mathrm{risk}}
% +\lambda_{\mathrm{SP}}
% \|\theta_R-\theta_R^{(0)}\|_2^2,
% \label{eq:total_loss}
% \end{equation}
% where $\mathcal{L}_{\mathrm{risk}}=\frac{1}{2N}\sum_{i=1}^{N}[(r_i^R)^2+(r_i^E)^2]$ penalizes large molecular risk scores. 
% The second regularization termconstrains the fine-tuned RNA encoder parameters $\theta_R$ toward their pretrained values $\theta_R^{(0)}$. We set $\lambda_r=10^{-3}$ and $\lambda_{\mathrm{SP}}=10^{-2}$.

\subsection{Data Collection and Preprocessing}

We obtained somatic mutation,  CNV, RNA sequencing (RNA-seq), and clinical data from TCGA~\cite{TCGA} for six cancer types: bladder urothelial carcinoma (BLCA), breast invasive carcinoma (BRCA), HNSC, LIHC, ovarian serous cystadenocarcinoma (OV), and uterine corpus endometrial carcinoma (UCEC). We included patients with matched molecular data and available age at diagnosis, tumor stage, and overall survival (OS) information.
%We collected somatic mutation, copy-number variation (CNV), RNA sequencing (RNA-seq), and clinical data from The Cancer Genome Atlas (TCGA)~\cite{TCGA}. Six cancer types were included: bladder urothelial carcinoma (BLCA), breast invasive carcinoma (BRCA), HNSC, LIHC, ovarian serous cystadenocarcinoma (OV), and uterine corpus endometrial carcinoma (UCEC). Patients were required to have matched molecular data, age at diagnosis, tumour stage, and overall survival (OS) information. %The six cancer-specific cohorts comprised 2,782 patients in total. 
Each cancer type was modeled and evaluated independently. An independent human papillomavirus (HPV)-negative HNSCC cohort from  CPTAC~\cite{CPTAC} was included for the subsequent HNSC case study. This cohort comprised 102 patients, including 39 observed deaths. It was used to evaluate the prognostic associations identified in TCGA-HNSC. Cohort characteristics are summarized in Table~\ref{tab:cohorts}.

% % Somatic mutations were annotated with clonal status~\cite{PyClone} and functional consequences. Mutation and CNV features were used to construct the 38-dimensional eTime representation,  as detailed in
% Section~\ref{etime}. RNA-seq expression values were log-transformed. The 2,000 genes with the highest median absolute deviation  were selected and standardized. Clinical variables included age and tumour stage. Age was standardized, and tumour stage was one-hot encoded.

Somatic mutation and CNV data were used to construct eTime as described in Section~\ref{etime}. RNA-seq expression values were log-transformed, and the 2,000 genes with the highest median absolute deviation (MAD) were selected and standardized. Clinical variables included age at diagnosis and tumor stage. Age was standardized, and tumor stage was one-hot encoded.

\vspace{-1em}
\begin{table}[h]
\centering
\caption{Characteristics of the six TCGA cancer cohorts and the independent CPTAC-HNSCC cohort.}
\label{tab:cohorts}
\begin{tabular*}{\textwidth}{@{\extracolsep{\fill}}lrrrrrrr@{}}
\toprule
 & BLCA & BRCA & HNSC & LIHC & OV & UCEC & CPTAC-HNSCC \\
\midrule
Patients     & 385 & 876 & 481 & 320 & 263 & 457 & 102 \\
Events       & 168 & 114 & 205 & 109 & 156 & 77  & 39 \\
Censored (\%) & 56.4 & 87.0 & 57.4 & 65.9 & 40.7 & 83.2 & 61.8 \\
\bottomrule
\end{tabular*}
\end{table}

\vspace{-2.5em}
\subsection{Evaluation Metrics and Experimental Setup}

We evaluated the survival prediction performance of EvoSurv using the C-index, which measures the agreement between predicted risk scores and observed survival outcomes. For a comparable pair of patients, the prediction is considered concordant if the patient with a shorter observed survival time and an observed event receives a higher predicted risk score. For right-censored data, pairs in which the patient with the shorter observed survival time is censored are excluded. The C-index is defined as
\begin{equation}
C =
\frac{
\sum_{i=1}^{N}\sum_{j=1}^{N}
\mathbb{I}(T_i<T_j,\delta_i=1)
\left[
\mathbb{I}(r_i>r_j)
+\frac{1}{2}\mathbb{I}(r_i=r_j)
\right]
}{
\sum_{i=1}^{N}\sum_{j=1}^{N}
\mathbb{I}(T_i<T_j,\delta_i=1)
},
\end{equation}
where $N$ denotes the number of patients, $T_i$ is the observed survival time of patient $i$, $\delta_i$ is the event indicator, $r_i$ is the predicted risk score, and $\mathbb{I}(\cdot)$ denotes the indicator function. The C-index ranges from 0 to 1, with 0.5 indicating random risk ranking and 1.0 indicating perfect concordance. Higher values indicate better risk discrimination. %The C-index ranges from 0 to 1, with higher values indicating better risk discrimination. A value of 0.5 corresponds to random risk ranking, while 1.0 indicates perfect concordance.

For each TCGA cancer cohort, patients were randomly divided into training (64\%), validation (16\%), and test (20\%) sets, stratified by survival event status. All data-dependent preprocessing parameters were fitted exclusively on the training set. The procedure was repeated using 50 random seeds (0--49), and test C-index values were reported as the mean and standard deviation across the 50 runs.
The 95\% confidence intervals (CI) of the mean test C-index were estimated using the corrected resampled $t$ method of Nadeau and Bengio~\cite{nadeau1999inference} to account for dependence across repeated data splits.
EvoSurv was compared with CoFormerSurv~\cite{CoFormerSurv}, FGCNSurv~\cite{FGCNSurv}, and AGGSurv~\cite{AGGSurv} using identical data splits. The baseline methods were implemented using their original code and recommended hyperparameters. Additional details on prognostic association testing, including out-of-fold risk estimation, nested Cox models, and statistical validation, are provided in Supplementary Methods (Section~S1).

To establish EvoSurv's predictive performance before investigating evolutionary features, we compared it with the three baseline methods across six TCGA cohorts using 50 random splits. %As shown in Table~\ref{tab:benchmark}, EvoSurv achieved the highest mean C-index in every cohort, with an overall mean of 0.661—relative improvements of 5.4\%, 7.6\%, and 8.0\% over the three baselines, respectively.

\vspace{-0.5em}
\section{Results}

\subsection{EvoSurv Achieves Strong Performance Across Six Cancer Cohorts}

% To evaluate the prognostic performance of EvoSurv, we compared it with three baseline methods (CoFormerSurv~\cite{CoFormerSurv}, FGCNSurv~\cite{FGCNSurv}, and AGGSurv~\cite{AGGSurv}) across six TCGA cancer cohorts (Table~\ref{tab:benchmark}). Performance was assessed over 50 random splits, with mean C-index, standard deviation, and 95\% confidence intervals reported in Table~\ref{tab:benchmark}. 

EvoSurv achieved the highest mean C-index among the evaluated methods across all six cancer cohorts, with an overall average of 0.663, representing relative improvements of 5.6\%, 7.8\%, and 8.2\% over CoFormerSurv, FGCNSurv, and AGGSurv, respectively (Table~\ref{tab:benchmark}).
% The corrected 95\% CIs reflect variability across random splits.
The results demonstrate the predictive performance of the integrated framework, while the contribution of eTime is examined separately below. 
EvoSurv achieved particularly strong predictive performance in UCEC and BRCA, with C-indices of 0.716 and 0.699, respectively, highlighting the potential benefits of multimodal integration. Consistent improvements were also observed in BLCA, LIHC, and OV, with relative gains of approximately 4.8\%--5.8\% over the best-performing baseline in each cancer. EvoSurv also achieved the highest mean C-index in HNSC, providing a basis for further investigation of the prognostic relevance of tumor evolutionary features in this cohort.

\begin{table}[h]
\centering
\caption{Survival prediction and modality ablation across six TCGA cohorts (mean C-index $\pm$ SD over 50 splits, with 95\% CIs in parentheses). The last column shows the cohort average. Best results are bold.}
\label{tab:benchmark}
\setlength{\tabcolsep}{2pt}
\renewcommand{\arraystretch}{1.1}
\begin{tabular*}{\textwidth}{@{\extracolsep{\fill}}lccccccc@{}}
\toprule
Method & BLCA & BRCA & HNSC & LIHC & OV & UCEC & Mean \\
\midrule
\multicolumn{8}{l}{\textit{Baseline methods}} \\
CoFormerSurv
& \shortstack[c]{$0.622 \pm 0.046$\\[-1pt] $(0.569, 0.674)$}
& \shortstack[c]{$0.648 \pm 0.070$\\[-1pt] $(0.567, 0.728)$}
& \shortstack[c]{$0.603 \pm 0.048$\\[-1pt] $(0.547, 0.658)$}
& \shortstack[c]{$0.628 \pm 0.059$\\[-1pt] $(0.559, 0.696)$}
& \shortstack[c]{$0.606 \pm 0.055$\\[-1pt] $(0.542, 0.669)$}
& \shortstack[c]{$0.659 \pm 0.059$\\[-1pt] $(0.591, 0.727)$}
& 0.628 \\
\addlinespace[3pt]
FGCNSurv
& \shortstack[c]{$0.620 \pm 0.041$\\[-1pt] $(0.572, 0.668)$}
& \shortstack[c]{$0.635 \pm 0.059$\\[-1pt] $(0.566, 0.703)$}
& \shortstack[c]{$0.607 \pm 0.039$\\[-1pt] $(0.562, 0.652)$}
& \shortstack[c]{$0.621 \pm 0.053$\\[-1pt] $(0.559, 0.682)$}
& \shortstack[c]{$0.603 \pm 0.054$\\[-1pt] $(0.540, 0.665)$}
& \shortstack[c]{$0.605 \pm 0.070$\\[-1pt] $(0.524, 0.686)$}
& 0.615 \\
\addlinespace[3pt]
AGGSurv
& \shortstack[c]{$0.613 \pm 0.045$\\[-1pt] $(0.561, 0.666)$}
& \shortstack[c]{$0.620 \pm 0.076$\\[-1pt] $(0.532, 0.707)$}
& \shortstack[c]{$0.592 \pm 0.044$\\[-1pt] $(0.541, 0.642)$}
& \shortstack[c]{$0.623 \pm 0.066$\\[-1pt] $(0.546, 0.699)$}
& \shortstack[c]{$0.559 \pm 0.053$\\[-1pt] $(0.497, 0.620)$}
& \shortstack[c]{$0.668 \pm 0.066$\\[-1pt] $(0.591, 0.745)$}
& 0.613 \\
\midrule
\multicolumn{8}{l}{\textit{Modality ablation}} \\
Clinical + RNA
& \shortstack[c]{$0.656 \pm 0.040$\\[-1pt] $(0.610, 0.702)$}
& \shortstack[c]{$0.663 \pm 0.057$\\[-1pt] $(0.597, 0.730)$}
& \shortstack[c]{$0.609 \pm 0.048$\\[-1pt] $(0.554, 0.664)$}
& \shortstack[c]{$0.653 \pm 0.048$\\[-1pt] $(0.597, 0.708)$}
& \shortstack[c]{$0.628 \pm 0.056$\\[-1pt] $(0.564, 0.693)$}
& \shortstack[c]{$0.698 \pm 0.083$\\[-1pt] $(0.602, 0.794)$}
& 0.651 \\
\addlinespace[3pt]
Clinical + eTime
& \shortstack[c]{$0.631 \pm 0.052$\\[-1pt] $(0.571, 0.691)$}
& \shortstack[c]{$0.685 \pm 0.039$\\[-1pt] $(0.640, 0.730)$}
& \shortstack[c]{$0.570 \pm 0.061$\\[-1pt] $(0.500, 0.641)$}
& \shortstack[c]{$0.621 \pm 0.052$\\[-1pt] $(0.561, 0.682)$}
& \shortstack[c]{$0.612 \pm 0.057$\\[-1pt] $(0.546, 0.677)$}
& \shortstack[c]{$0.710 \pm 0.074$\\[-1pt] $(0.625, 0.796)$}
& 0.638 \\
\midrule
Full (EvoSurv)
& \shortstack[c]{$\mathbf{0.658 \pm 0.039}$\\[-1pt] $(0.613, 0.703)$}
& \shortstack[c]{$\mathbf{0.699 \pm 0.051}$\\[-1pt] $(0.640, 0.759)$}
& \shortstack[c]{$\mathbf{0.612 \pm 0.046}$\\[-1pt] $(0.560, 0.665)$}
& \shortstack[c]{$\mathbf{0.660 \pm 0.059}$\\[-1pt] $(0.591, 0.728)$}
& \shortstack[c]{$\mathbf{0.635 \pm 0.056}$\\[-1pt] $(0.570, 0.699)$}
& \shortstack[c]{$\mathbf{0.716 \pm 0.069}$\\[-1pt] $(0.636, 0.796)$}
& $\mathbf{0.663}$ \\
\bottomrule
\end{tabular*}
\end{table}

% \paragraph{Modality Ablation Study.}
\subsubsection{Modality Ablation Study}

% To assess the contributions of RNA and eTime under a fixed clinical risk component, we compared Clinical + RNA, Clinical + eTime, and the full model. The full model achieved the highest mean C-index across all six cohorts (0.663), compared with 0.651 and 0.638, respectively (Table~\ref{tab:benchmark}). Adding eTime yielded the largest gains in BRCA (+0.036) and UCEC (+0.018). 
% In addition, RNA + eTime without the clinical risk component achieved a mean C-index of 0.621 (Supplementary Table~S1). These results suggest complementary predictive value from eTime, with varying gains across cancers.

To assess the contributions of RNA and eTime under a fixed clinical risk component, we compared Clinical + RNA, Clinical + eTime, and the full model. The full model outperformed both reduced configurations across all six cohorts, with the largest gains from adding eTime observed in BRCA (+0.036) and UCEC (+0.018). In addition, RNA + eTime without the clinical risk component achieved a mean C-index of 0.621 (Supplementary Table~S1).  These results suggest complementary predictive value from eTime, with varying gains across cancers.

% \vspace{-1em}
%\subsection{Complementary Prognostic Value of eTime}
\subsection{Does eTime provide additional prognostic signals?}

\label{sec:complementary}

To assess the prognostic value of eTime, we examined whether its evolutionary features contain survival-associated information and provide additional prognostic value beyond clinical and transcriptomic predictors.

% \vspace{-2em}
%\subsubsection{eTime Contains Survival-Associated Information}
\subsubsection{eTime Contains Survival-Associated Information}

% We compared the eTime-only survival model with a shuffled control across 50 random splits to assess its standalone predictive information (see Supplementary Methods).

We first compared an eTime-only survival model across 50 random data splits. As a reference, we applied the same training and evaluation procedure after randomly shuffling eTime profiles across patients, disrupting their correspondence with survival outcomes. This provided a matched baseline for assessing the predictive information contained in evolutionary features.

Across all six cancer cohorts, eTime achieved higher mean test C-indices than the shuffled baseline, ranging from 0.529 to 0.645 compared with 0.489 to 0.509 (Table~\ref{tab:etime_alone}).
The greatest differences were observed in UCEC ($\Delta C=0.156$), LIHC ($\Delta C=0.071$), and OV ($\Delta C=0.063$). These results indicate that eTime contains survival-associated information, although its standalone discrimination varied across cancer types.%with varying predictive contributions across cancer types.

% \begin{table}[h]
% \centering
% \caption{Test C-index of the eTime-only model (mean over 50 splits). Shuffled: eTime profiles shuffled across patients, with identical training and evaluation.}
% \label{tab:etime_alone}
% \begin{tabular*}{\textwidth}{@{\extracolsep{\fill}}lcccccc@{}}
% \toprule
%  & BLCA & BRCA & HNSC & LIHC & OV & UCEC \\
% \midrule
% eTime      & 0.531 & 0.539 & 0.529 & 0.580 & 0.556 & 0.645 \\
% Shuffled   & 0.502 & 0.498 & 0.495 & 0.509 & 0.493 & 0.489 \\
% $\Delta C$ & +0.030 & +0.041 & +0.034 & +0.070 & +0.063 & +0.156 \\
% \bottomrule
% \end{tabular*}
% \end{table}
% \vspace{-1em}
\begin{table}[h]
\centering
\caption{Prognostic performance of eTime and shuffled features across 50 random splits (mean C-index $\pm$ SD; 95\% CIs in parentheses). $\Delta C$ denotes the performance difference.}
\label{tab:etime_alone}
% \normalsize
% \scriptsize
% \renewcommand{\arraystretch}{1.15}
\begin{tabular*}{\textwidth}{@{\extracolsep{\fill}}lccccccc@{}}
\toprule
Method & BLCA & BRCA & HNSC & LIHC & OV & UCEC & Overall \\
\midrule

eTime
& \shortstack[c]{0.531 $\pm$ 0.045\\[-1pt] (0.480, 0.583)}
& \shortstack[c]{0.539 $\pm$ 0.070\\[-1pt] (0.457, 0.620)}
& \shortstack[c]{0.529 $\pm$ 0.052\\[-1pt] (0.469, 0.589)}
& \shortstack[c]{0.580 $\pm$ 0.056\\[-1pt] (0.515, 0.645)}
& \shortstack[c]{0.556 $\pm$ 0.053\\[-1pt] (0.494, 0.617)}
& \shortstack[c]{0.645 $\pm$ 0.069\\[-1pt] (0.565, 0.724)}
& \shortstack[c]{0.563 $\pm$ 0.024\\[-1pt] (0.535, 0.591)} \\
\addlinespace[3pt]

Shuffled
& \shortstack[c]{0.502 $\pm$ 0.054\\[-1pt] (0.439, 0.564)}
& \shortstack[c]{0.498 $\pm$ 0.072\\[-1pt] (0.415, 0.581)}
& \shortstack[c]{0.495 $\pm$ 0.047\\[-1pt] (0.441, 0.550)}
& \shortstack[c]{0.509 $\pm$ 0.074\\[-1pt] (0.423, 0.595)}
& \shortstack[c]{0.493 $\pm$ 0.059\\[-1pt] (0.425, 0.561)}
& \shortstack[c]{0.489 $\pm$ 0.088\\[-1pt] (0.387, 0.591)}
& \shortstack[c]{0.498 $\pm$ 0.026\\[-1pt] (0.468, 0.527)} \\

\midrule
% $\Delta C$
% & +0.030
% & +0.040
% & +0.033
% & +0.070
% & +0.063
% & +0.156
% & +0.065 \\

$\Delta C$
& +0.029
& +0.041
& +0.034
& +0.071
& +0.063
& +0.156
& +0.065 \\

\bottomrule
\end{tabular*}
\end{table}

\vspace{-2em}
%\subsubsection{Complementary Value beyond Clinical Factors and RNA.}

\subsubsection{eTime Retains an Association After Adjustment for Clinical and RNA Predictors.}
We next examined whether the prognostic information captured by eTime persisted after adjustment for established predictors. We used L2-regularized Cox models with five-fold cross-validation repeated five times 
to generate separate eTime and RNA risk scores. Each patient's out-of-fold score was generated by models trained without that patient. We then evaluated the eTime risk score in nested Cox proportional hazards models adjusted for age and tumor stage, the RNA risk score, or both. Hazard ratios (HR) were estimated per standard deviation of the eTime risk score. We used one-sided likelihood-ratio tests and combined evidence across cancer cohorts using Fisher's method~\cite{combine_p}.

%The 38-dimensional eTime representation and the RNA expression profiles were separately summarized into scalar risk scores by L2-regularized Cox models, using five-fold cross-validation repeated five times. Each patient was scored only by models trained without that patient, so that the resulting out-of-fold scores are free of optimistic bias in the subsequent association analysis.

% We evaluated the eTime risk score using nested Cox proportional hazards models adjusted for clinical variables (age and tumor stage), the RNA risk score, or both. Because the eTime score is a risk score with an expected positive association, one-sided likelihood-ratio tests were used. Hazard ratios were reported per standard deviation of the eTime risk score, and evidence across cancer cohorts was combined using Fisher's method.

% After adjustment for clinical variables, eTime was associated with survival in HNSC (HR $=1.23$, $p=0.003$) and LIHC (HR $=1.28$, $p=0.006$), as shown in Table~\ref{tab:etime_incremental}. The combined evidence across cohorts was significant (Fisher's $p=3.1\times10^{-4}$).
As shown in Table~\ref{tab:etime_incremental}, after adjustment for clinical variables, higher eTime risk scores were associated with a higher hazard of death in HNSC (HR $=1.23$, $p=0.003$) and LIHC (HR $=1.28$, $p=0.006$), with significant combined evidence across cohorts (Fisher's $p=3.1\times10^{-4}$).
After adjustment for the RNA risk score, positive associations between the eTime risk score and survival were observed in HNSC (HR = 1.17, \(p=0.018\)), LIHC (HR = 1.19, \(p=0.034\)), and UCEC (HR = 1.24, \(p=0.047\)), with a combined Fisher's $p=0.003$.
% . Evidence combined across all six cancer cohorts was significant (Fisher’s \(p=0.003\)). 
Importantly, after simultaneous adjustment for clinical variables and the RNA risk score, eTime remained associated with survival in HNSC (HR $=1.20$, $p=0.009$) and LIHC (HR $=1.22$, $p=0.017$). Evidence combined across all six cohorts was significant (Fisher’s \(p=0.009\)). %after simultaneous adjustment for clinical and transcriptomic predictors, and 
These findings indicate that eTime captures survival-associated information that is not fully represented by clinical and RNA risk scores, particularly in HNSC and LIHC.%the complementary prognostic relevance of evolutionary information, particularly in HNSC and LIHC.

\vspace{-1em}
\begin{table}[h]
\centering
\caption{Association of the eTime risk score with survival after adjustment for clinical and RNA predictors. %HR is per standard deviation of the out-of-fold eTime risk score, adjusted for stage and age (Clinical), the out-of-fold RNA risk score (RNA), or both (Clinical + RNA). $p$-values are from one-sided likelihood-ratio tests; confidence intervals are two-sided.
HRs are reported per standard deviation of the out-of-fold eTime risk score. Cohort-specific \(p\)-values are from one-sided likelihood-ratio tests. Fisher’s \(p\)-values combine evidence across the six cancer cohorts.}
\label{tab:etime_incremental}
% \footnotesize
\setlength{\tabcolsep}{2pt}
\begin{tabular*}{\textwidth}{@{\extracolsep{\fill}}lcccccc@{}}
\toprule
 & \multicolumn{2}{c}{Clinical} & \multicolumn{2}{c}{RNA} & \multicolumn{2}{c}{Clinical + RNA} \\
\cmidrule(lr){2-3}\cmidrule(lr){4-5}\cmidrule(lr){6-7}
Cancer & HR (95\% CI) & $p$ & HR (95\% CI) & $p$ & HR (95\% CI) & $p$ \\
\midrule
BLCA & 1.10 (0.93--1.29) & 0.137 & 1.02 (0.88--1.20) & 0.386 & 1.05 (0.89--1.23) & 0.287 \\
BRCA & 0.96 (0.80--1.16) & 0.671 & 1.05 (0.88--1.25) & 0.305 & 0.93 (0.77--1.13) & 0.764 \\
HNSC & \textbf{1.23 (1.06--1.43)} & \textbf{0.003} & \textbf{1.17 (1.01--1.36)} & \textbf{0.018} & \textbf{1.20 (1.03--1.40)} & \textbf{0.009} \\
LIHC & \textbf{1.28 (1.05--1.56)} & \textbf{0.006} & \textbf{1.19 (0.98--1.43)} & \textbf{0.034} & \textbf{1.22 (1.01--1.48)} & \textbf{0.017} \\
OV   & 1.13 (0.97--1.31) & 0.064 & 1.10 (0.94--1.27) & 0.113 & 1.09 (0.93--1.27) & 0.135 \\
UCEC & 1.14 (0.90--1.44) & 0.134 & \textbf{1.24 (0.96--1.58)} & \textbf{0.047} & 1.04 (0.80--1.35) & 0.380 \\
\midrule
Fisher's combined $p$ & & $\mathbf{3.1\times10^{-4}}$ & & \textbf{0.003} & & \textbf{0.009} \\
\bottomrule
\end{tabular*}
\end{table}

% \vspace{-1.8em}
% \subsection{Contribution and Attribution of Evolutionary Features}

%\subsection{Evolutionary Feature Associations}

\subsection{Which Evolutionary Features Are Prognostically Associated, and Do They Vary by Cancer Type?}
\label{sec:Contribution}

\begin{figure}[h]
    \centering
    \includegraphics[width= \linewidth]{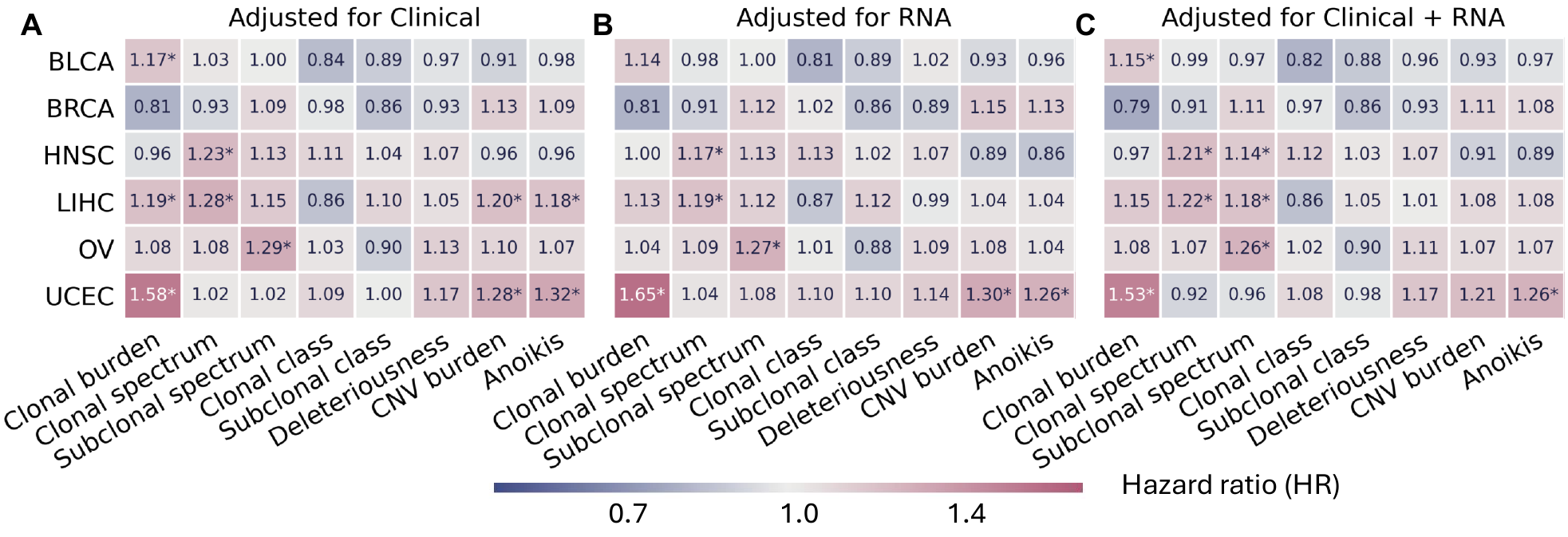}
    \caption{\textbf{Prognostic associations of evolutionary features across six cancers.} Heatmaps show hazard ratios (HRs) after adjustment for (A) clinical variables (age and stage), (B) RNA risk scores, and (C) both. Asterisks indicate statistical significance ($p < 0.05$).}
    \label{fig:heatmap}
\end{figure}

%To further investigate the prognostic relevance of tumor evolutionary features, we evaluated the associations of the eight previously defined eTime feature groups with overall survival. Three adjustment settings were considered: clinical variables (age and tumor stage), RNA risk scores, and both (Fig.~\ref{fig:heatmap}). 
To identify the eTime features associated with survival, we evaluated  eight feature groups using Cox models under three adjustment settings: clinical variables (age and tumor stage), RNA risk score, and both (Fig.~\ref{fig:heatmap}). The observed associations varied across cancer types and adjustment settings, indicating that the prognostic relevance of evolutionary features is cancer-specific. %potential overlap in prognostic information between evolutionary, clinical, and transcriptomic features. 
% We focus below on UCEC, LIHC, and HNSC, which showed distinct feature patterns and in which the overall eTime score was associated with survival after adjustment for clinical and transcriptomic predictors. In OV, the subclonal mutation spectrum showed a relatively consistent association with survival. UCEC, HNSC, and LIHC exhibited distinct patterns of prognostic associations. We therefore focus on these three cancer types below.
In OV, the subclonal mutational spectrum remained associated with survival under all three settings. We further focused on HNSC and LIHC, where the overall eTime risk score remained associated with survival after adjustment for clinical and transcriptomic predictors, and UCEC, where specific evolutionary feature groups showed prognostic associations.

%\vspace{-1em}

In UCEC, the burden of clonal mutations showed  association with increased hazard in all three adjustment settings (HRs$=1.58$, $1.65$, and $1.53$ after adjustment for Clinical, RNA, and both). CNV- and anoikis-related features also showed associations with increased hazard, while the anoikis feature group remained associated after joint adjustment (HR $= 1.26$). %with overall survival. The hazard ratios (HRs) were 1.58, 1.65, and 1.53 after adjustment for Clinical, RNA, and Clinical + RNA, respectively. CNV and anoikis-related features also showed positive associations, with the anoikis feature group retaining an association after joint adjustment (HR = 1.26). These findings may related to the molecular heterogeneity of UCEC. TCGA classified endometrial carcinoma into four molecular subtypes: POLE ultramutated, microsatellite instability (MSI), copy-number low, and copy-number high. These subtypes differ in mutation burden, chromosomal alterations, and clinical outcomes~\cite{UCEC_subtype}. Notably, POLE-ultramutated tumors generally have favorable outcomes despite their high mutation burden. Previous studies have also linked anoikis-related gene expression to UCEC prognosis~\cite{UCEC_anoikis}. 
These findings may be related to the molecular heterogeneity of UCEC. TCGA classified endometrial carcinoma into four molecular subtypes: POLE ultramutated, microsatellite instability (MSI), copy-number low, and copy-number high. These subtypes differ in mutation burden, chromosomal alterations, and clinical outcomes~\cite{UCEC_subtype}. Notably, POLE-ultramutated tumors generally have favorable outcomes despite their high mutation burden. Previous studies have also linked anoikis-related gene expression to UCEC prognosis~\cite{UCEC_anoikis}. Our findings extend these observations to genomic alterations in anoikis-related genes but do not establish altered anoikis function.

%%%%%%%%% Here
In LIHC, clonal and subclonal mutation spectra remained associated with increased hazard after joint adjustment. The CNV-related association weakened after adjusting for RNA risk (HR: 1.20 to 1.08), suggesting that some CNV-related prognostic information may overlap with RNA expression. Overall, these findings suggest that mutation-spectrum and CNV features capture distinct aspects of prognostic information, consistent with previous LIHC multi-omics studies \cite{LIHC}. %Consistent with previous LIHC multi-omics studies [5], these findings suggest that mutation spectra and CNV-related features capture different aspects of prognostic information.

HNSC showed a distinct pattern: the clonal mutational spectrum remained associated with survival across all three adjustment settings (HRs = 1.23, 1.17, and 1.21, respectively), whereas the subclonal mutational spectrum showed weaker associations. This pattern is consistent with the association between the overall eTime risk score and survival after joint adjustment (HR = 1.20, \(p = 0.009\)).
The HNSC mutational profiles are shaped by multiple mutagenic processes, including HPV-associated processes, tobacco-related mutagenesis, and APOBEC activity~\cite{tcga2015hnsc,tobacco,APOBEC}. 
% HPV and tobacco-associated processes~\cite{tcga2015hnsc}, %, which exhibit distinct genomic alteration profiles~\cite{tcga2015hnsc}, 
% as well as tobacco-related mutagenesis and APOBEC activity~\cite{tobacco,APOBEC}.
Clonal mutations generally represent events acquired before or during major clonal expansions~\cite{hnsc_clonal}, and therefore their mutational spectrum may retain information on mutagenic processes active during earlier stages of tumor evolution. In contrast, RNA expression reflects the molecular state of the tumor at the time of sampling. This distinction may help explain why the mutational composition provides additional information beyond the transcriptome. 
% To study the specific nucleotide substitution types contribute to this association, 
To identify the specific substitution types contributing to this association, we next analyzed individual substitution classes.

%\subsection{HNSC Case Study: Prognostic and Biological Insights}
\subsection{HNSC Case Study: T>G Substitutions Show a Robust Prognostic Association }
\label{sec:casestudy}

\begin{figure}
    \centering
    \includegraphics[width=\linewidth]{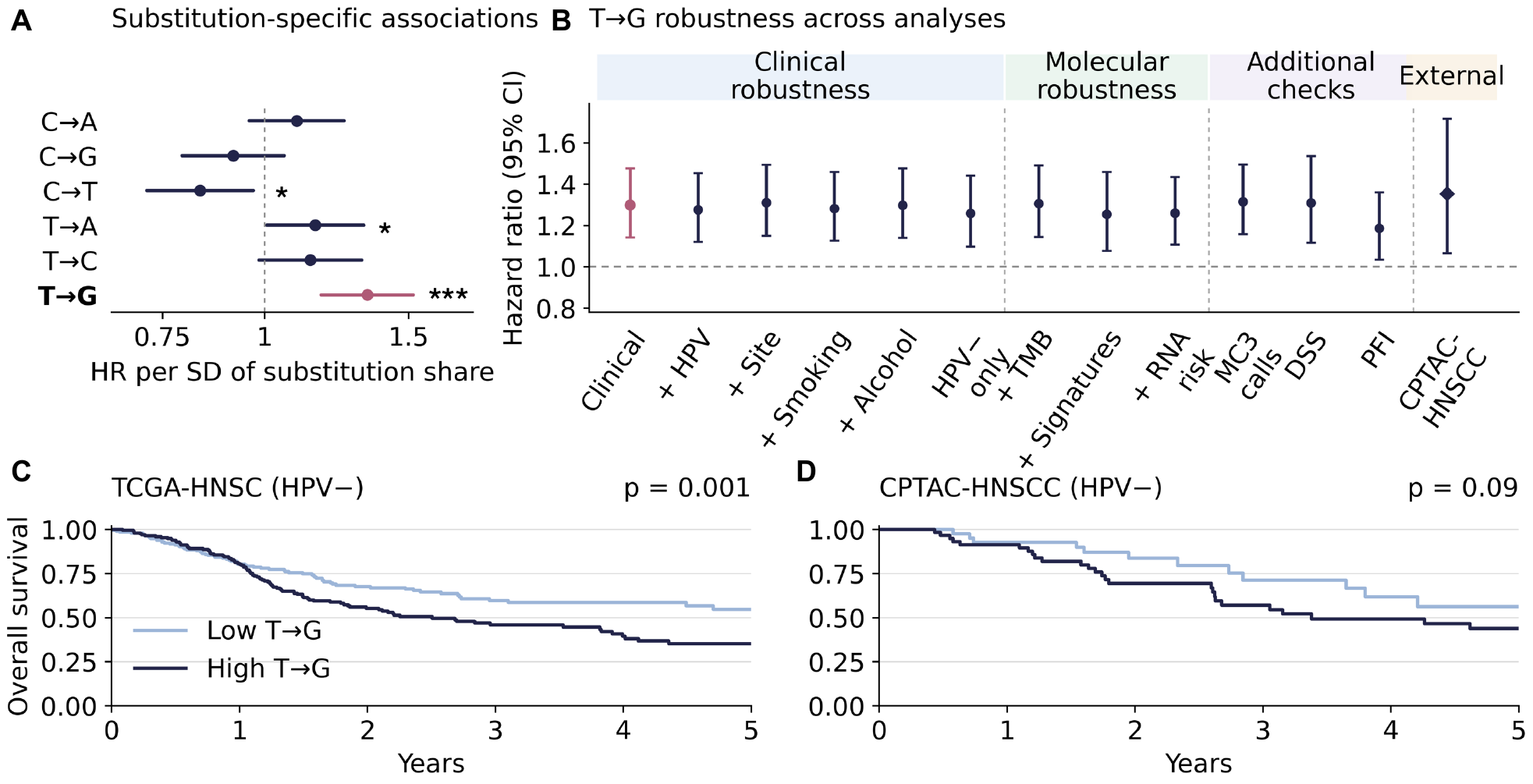}
\caption{\textbf{Prognostic significance and robustness of T$>$G substitutions in HNSC.}
(A) HRs and 95\% CIs for six substitution types. Asterisks indicate significance (*$p<0.05$, ***$p<0.001$).
(B) Robustness of T$>$G associations across clinical, molecular, and endpoint analyses, including external validation. All associations were significant ($p<0.05$).
(C,D) Kaplan--Meier curves stratified by T$>$G burden in HPV-negative TCGA-HNSC and CPTAC-HNSCC, respectively.}
    \label{fig:hnsc}
\end{figure}

To identify specific base substitution types associated with prognosis in HNSC, we evaluated the associations of six substitution classes with overall survival (Fig.~\ref{fig:hnsc}A). T$>$G showed the strongest positive association (HR = 1.34), 
whereas C$>$T showed a significant negative association and C$>$G exhibited a similar but nonsignificant negative trend. C$>$T and C$>$G substitutions in specific trinucleotide contexts characterize the APOBEC-related mutational signatures SBS2 and SBS13, respectively. APOBEC activity is also more prominent in HPV-positive head and neck cancers~\cite{APOBEC}.
% whereas C$>$T and C$>$G showed negative associations. C$>$T and C$>$G are characteristic substitutions of the APOBEC-related mutational signatures SBS2 and SBS13, respectively. APOBEC activity is also more prominent in HPV-positive head and neck cancers~\cite{APOBEC}. 
Notably, HPV-positive oropharyngeal cancer patients generally have better survival outcomes~\cite{hpv}. Thus, the negative associations of C$>$T and C$>$G may partly reflect HPV-related mutational patterns and survival differences. These observations motivated us to further examine the influence of HPV status on the prognostic association of T$>$G.

We  assessed the robustness of the T$>$G prognostic association against clinical factors, molecular characteristics, mutation-calling procedures, and survival endpoints (Fig.~\ref{fig:hnsc}B). Beyond age and tumor stage, we separately adjusted for HPV status, anatomical site, smoking, and alcohol consumption. The association remained consistent, with HRs ranging from 1.27 to 1.31. It also persisted in HPV-negative patients (HR = 1.26), indicating that the prognostic signal was not solely explained by HPV-related differences. We further adjusted for tumor mutational burden (TMB) and mutational signatures to account for overall mutation load and known mutagenic processes. The association remained stable, including after adjustment for the RNA risk score, supporting its complementary prognostic information. Consistent effect directions were also observed using the TCGA Pan-Cancer MC3 mutation call set and alternative survival endpoints, including disease-specific survival (DSS) and progression-free interval (PFI).

The association was further supported by the independent CPTAC-HNSCC cohort, where T$>$G was associated with increased mortality risk (HR = 1.35, Fig.~\ref{fig:hnsc}B). We also examined survival stratification by T$>$G proportion among HPV-negative patients. Kaplan--Meier curves showed poorer overall survival in the high-T$>$G group in TCGA-HNSC (adjusted Cox test, $p = 0.001$, Fig.~\ref{fig:hnsc}C). A similar survival trend was observed in CPTAC-HNSCC ($p = 0.09$, Fig.~\ref{fig:hnsc}D). %Overall, the continuous association and survival stratification analyses support the prognostic relevance of T$>$G in HNSC.
Overall, these analyses support T>G as a robust HNSC prognostic signal that persists beyond several clinical and molecular predictors, including the RNA risk score. This result provides a specific example of the complementary prognostic information captured by tumor evolutionary features.

\section{Discussion}

We developed eTime to characterize tumor evolutionary heterogeneity through clonal and subclonal mutation burden, mutational spectra, functional consequences, copy-number alterations, and anoikis-related genomic changes. Integrating eTime with clinical and transcriptomic information, EvoSurv achieved the highest mean C-index in all six cancer types. More importantly, multivariable Cox analyses showed that the eTime risk score remained associated with survival after adjustment for clinical variables and RNA-derived risk scores, particularly in HNSC and LIHC. %More importantly, multivariable Cox regression and incremental model comparisons revealed complementary prognostic information beyond clinical variables and RNA-derived risk scores, particularly in HNSC and LIHC. 
%The informative features differed across cancers: clonal burden was associated with increased mortality in UCEC across all adjustment settings, mutational spectra showed consistent prognostic associations in HNSC and OV, and the CNV association in LIHC weakened after RNA adjustment, suggesting overlapping prognostic information with transcriptomic features. 
 The associated evolutionary features varied across cancer types. Clonal mutation burden was associated with increased hazard in UCEC, mutation-spectrum features showed prognostic associations in HNSC and LIHC, and the subclonal mutation spectrum showed a relatively consistent association in OV. In LIHC, the CNV-related association weakened after adjustment for the RNA risk score. These findings indicate that evolutionary features capture distinct aspects of survival-associated heterogeneity, with contributions that vary across cancer types.

In HNSC, the clonal mutational spectrum showed consistent prognostic associations across all adjustment settings, motivating further analysis of individual substitution classes. A higher proportion of T$>$G substitutions was associated with increased mortality risk after adjustment for clinical and molecular covariates, including HPV status, with further support from the independent CPTAC-HNSCC cohort. Given the roles of HPV infection, APOBEC activity, and tobacco exposure in shaping the HNSC mutational landscape, the origins and biological implications of this association warrant further investigation. Trinucleotide mutational signature and pathway analyses could help identify the underlying mutational processes and associated biological pathways. Further functional validation may clarify their relevance to tumor progression and potential value for prognostic stratification in HNSC.

\section*{Acknowledgments}
Some of the icons in Figure 1 were generated using generative AI and reviewed by the authors. These icons are for illustrative purposes only.

\vspace{-1em}
\bibliographystyle{splncs04}
\bibliography{ref}
\end{document}

\begin{center}
    {\Large\bfseries Supplementary Material}\\[6pt]
    {\large EvoSurv: Evolution-Aware Multimodal Survival Modeling Reveals Complementary Prognostic Signals}
\end{center}

\renewcommand{\thesection}{S\arabic{section}}

\section{Prognostic and Statistical Analysis}
\label{supp:prognostic}
To assess the standalone prognostic value of eTime, we first trained an eTime-only survival model and compared it with a shuffled control, in which eTime profiles were randomly permuted across patients. Both models followed the same training and evaluation procedure over 50 random splits, and their predictive performance was compared using the mean test C-index.

To evaluate the incremental prognostic value of eTime, we generated out-of-fold (OOF) risk scores for RNA and eTime using L2-regularized Cox models with five-fold cross-validation repeated five times. Each patient was scored only by models trained without that patient. We then constructed nested Cox proportional hazards models, using clinical variables (age and tumor stage), RNA risk scores, or both as baseline predictors, and adding the eTime risk score:
\begin{equation}
h_i(t) = h_0(t)
\exp\left(
\boldsymbol{\beta}^{\top}\mathbf{z}_i
+ \gamma e_i
\right),
\end{equation}
where $\mathbf{z}_i$ represents the baseline predictors, $e_i$ is the standardized eTime risk score, and $h_0(t)$ is the baseline hazard. The hazard ratio is defined as $\mathrm{HR}=\exp(\gamma)$, representing the relative change in hazard per standard deviation increase in the eTime risk score. 
Incremental prognostic value was assessed using the likelihood-ratio statistic $LR = 2(\ell_{\mathrm{full}}-\ell_{\mathrm{base}})$, where $\ell_{\mathrm{full}}$ and $\ell_{\mathrm{base}}$ denote the maximized log-likelihoods of the full and baseline models, respectively. Because the risk scores were constructed with a prespecified positive risk direction, one-sided tests were used to assess positive prognostic associations.  The eight eTime feature groups were evaluated separately using the same adjustment strategy.

For the HNSC case study, we assessed substitution-specific prognostic associations using Cox models adjusted for age and tumor stage, reporting HRs and 95\% confidence intervals per standard deviation. We further evaluated the robustness of T$>$G across clinical and molecular adjustments, HPV-negative patients, alternative mutation data and survival endpoints. The independent CPTAC-HNSCC cohort was used for external validation with prespecified analyses and a one-sided test.

For Kaplan--Meier (KM) analysis, HPV-negative TCGA-HNSC patients were divided into high- and low-T$>$G groups using the median T$>$G substitution proportion. The same cutoff was applied to the CPTAC-HNSCC cohort. KM curves were used to visualize overall survival differences, and group associations were assessed using Cox models adjusted for age and tumor stage.

\newpage
\section{Modality Ablation Results Across Six TCGA Cohorts.}
\label{supp:ablation}

\begin{table}[h]
\renewcommand{\thetable}{S\arabic{table}}
\centering
\caption{Modality ablation results across six TCGA cohorts (mean C-index $\pm$ SD, with approximate 95\% corrected resampled CIs in parentheses). Best results are bold.}
\label{tab:supp_modality}
\setlength{\tabcolsep}{3pt}
\renewcommand{\arraystretch}{1.1}
\begin{tabular*}{\textwidth}{@{\extracolsep{\fill}}lcccc@{}}
\toprule
Cancer & Clinical + RNA & Clinical + eTime & RNA + eTime & Full (EvoSurv) \\
\midrule
BLCA
& \shortstack[c]{$0.656 \pm 0.040$\\[-1pt] $(0.610, 0.702)$}
& \shortstack[c]{$0.631 \pm 0.052$\\[-1pt] $(0.571, 0.691)$}
& \shortstack[c]{$0.621 \pm 0.042$\\[-1pt] $(0.572, 0.670)$}
& \shortstack[c]{$\mathbf{0.658 \pm 0.039}$\\[-1pt] $(0.613, 0.703)$} \\
\addlinespace[3pt]
BRCA
& \shortstack[c]{$0.663 \pm 0.057$\\[-1pt] $(0.597, 0.730)$}
& \shortstack[c]{$0.685 \pm 0.039$\\[-1pt] $(0.640, 0.730)$}
& \shortstack[c]{$0.607 \pm 0.066$\\[-1pt] $(0.531, 0.683)$}
& \shortstack[c]{$\mathbf{0.699 \pm 0.051}$\\[-1pt] $(0.640, 0.759)$} \\
\addlinespace[3pt]
HNSC
& \shortstack[c]{$0.609 \pm 0.048$\\[-1pt] $(0.554, 0.664)$}
& \shortstack[c]{$0.570 \pm 0.061$\\[-1pt] $(0.500, 0.641)$}
& \shortstack[c]{$0.589 \pm 0.042$\\[-1pt] $(0.541, 0.638)$}
& \shortstack[c]{$\mathbf{0.612 \pm 0.046}$\\[-1pt] $(0.560, 0.665)$} \\
\addlinespace[3pt]
LIHC
& \shortstack[c]{$0.653 \pm 0.048$\\[-1pt] $(0.597, 0.708)$}
& \shortstack[c]{$0.621 \pm 0.052$\\[-1pt] $(0.561, 0.682)$}
& \shortstack[c]{$0.649 \pm 0.061$\\[-1pt] $(0.578, 0.720)$}
& \shortstack[c]{$\mathbf{0.660 \pm 0.059}$\\[-1pt] $(0.591, 0.728)$} \\
\addlinespace[3pt]
OV
& \shortstack[c]{$0.628 \pm 0.056$\\[-1pt] $(0.564, 0.693)$}
& \shortstack[c]{$0.612 \pm 0.057$\\[-1pt] $(0.546, 0.677)$}
& \shortstack[c]{$0.594 \pm 0.060$\\[-1pt] $(0.525, 0.662)$}
& \shortstack[c]{$\mathbf{0.635 \pm 0.056}$\\[-1pt] $(0.570, 0.699)$} \\
\addlinespace[3pt]
UCEC
& \shortstack[c]{$0.698 \pm 0.083$\\[-1pt] $(0.602, 0.794)$}
& \shortstack[c]{$0.710 \pm 0.074$\\[-1pt] $(0.625, 0.796)$}
& \shortstack[c]{$0.665 \pm 0.076$\\[-1pt] $(0.577, 0.753)$}
& \shortstack[c]{$\mathbf{0.716 \pm 0.069}$\\[-1pt] $(0.636, 0.796)$} \\
\midrule
Mean & 0.651 & 0.638 & 0.621 & $\mathbf{0.663}$ \\
\bottomrule
\end{tabular*}
\end{table}